\documentclass[a4paper,11pt]{article}
\usepackage{amsmath}
\usepackage{color}
\usepackage{graphicx}
\usepackage{amssymb}
\usepackage{float}
\usepackage{scrtime}
\usepackage{fancyhdr}
\usepackage{subfigure}
\usepackage{authblk}
\usepackage{lipsum}
\usepackage{rotating}
\usepackage{floatpag}
\usepackage{varioref}
\usepackage{slashed}
\usepackage{enumerate}
\usepackage[toc,page]{appendix}

\usepackage[colorlinks=true
,urlcolor=blue
,anchorcolor=blue
,citecolor=blue
,filecolor=blue
,linkcolor=black
,menucolor=blue
,pagecolor=blue
,linktocpage=true
]{hyperref}

\usepackage[numbers,sort&compress]{natbib}

\setlipsumdefault{131}

\def\be{\begin{equation}}
\def\ee{\end{equation}}
\def\bea{\begin{eqnarray}}
\def\eea{\end{eqnarray}}
\def\ls{\mathrel{\lower4pt\vbox{\lineskip=0pt\baselineskip=0pt
           \hbox{$<$}\hbox{$\sim$}}}}
\def\gs{\mathrel{\lower4pt\vbox{\lineskip=0pt\baselineskip=0pt
           \hbox{$>$}\hbox{$\sim$}}}}

\newcommand{\gsim}{\lower.7ex\hbox{$\;\stackrel{\textstyle>}{\sim}\;$}}
\newcommand{\lsim}{\lower.7ex\hbox{$\;\stackrel{\textstyle<}{\sim}\;$}}

\newcommand{\neu}[1]{\ensuremath{\tilde{\chi}_{#1}^0}}

\graphicspath{{./figures/}}

\begin{document}

\floatpagestyle{plain}

\pagenumbering{roman}

\renewcommand{\headrulewidth}{0pt}
\fancyfoot{}

\title{\huge \bf{Non-thermal Dark Matter: A Selective Aper\c{c}u}}

\author {Kuver Sinha}

\affil[$^\ddag$]{\em Department of Physics, Syracuse University, Syracuse, NY 13244, USA\enspace\enspace\enspace\enspace\enspace}

\maketitle
\thispagestyle{fancy}

\begin{abstract}\normalsize\parindent 0pt\parskip 5pt

I provide a review of some recent work on non-thermal dark matter.

\end{abstract}

\clearpage
\newpage

\pagenumbering{arabic}

\tableofcontents 

\section{Introduction: The Thermal Discontent}

The thermal WIMP paradigm holds that Weakly Interacting Massive Particles (WIMP)s can explain the dark matter (DM) relic abundance, as measured by cosmic microwave background experiments~\cite{WMAP}, via thermal freeze-out of annihilation in the early universe. It is predictive, minimal, and independent of the prior thermal history of the universe.

From the point of view of particle physics, however, the picture is in a sense \textit{too} minimal and predictive, and rules out vast regions of parameter space in well-motivated extensions of the Standard Model (SM). For most of these notes, we will be concerned with supersymmetry, and in particular our primary example will be the $R$-parity conserving Minimal Supersymmetric Standard Model (MSSM). We will make no value judgement of the MSSM with regard to its other discontents, such as naturalness and its unease with the observed Higgs mass. We will use it merely as an example of the kind of hard constraints that DM physics can put on a particle theory model.

The fact that the MSSM generically \textit{does not} prefer a thermal WIMP is well known. The  lightest supersymmetric particle (LSP), typically the lightest neutralino $\neu{1}$ protected by $R$-parity, is the canonical choice for the WIMP. However, assuming a thermal history, the WIMP gives the observed relic density only in fine-tuned regions of parameter space. For Wino or Higgsino DM, thermal history predicts a WIMP mass in the region of $2.5$ TeV and $1$ TeV, respectively, due to the fact that these candidates annihilate too efficiently for smaller mass. While there is nothing wrong with having such massive candidates, this takes such candidates out of contention for observation at colliders or DM experiments in the near future. For  Bino DM, the problem is that it over-produces DM unless its annihilation cross section is bolstered with coannihilation effects or resonances. These effects are generally fine-tuned \cite{ksaa}. 

Another possibility is a well tempered neutralino, which contains just the right relative projections of $\neu{1}$ along Bino and Wino/Higgsino directions, to give the observed relic density. This option is quite non-generic in parameter space, as may be expected given that a priori the projections can take any value they want. Moreover, some well tempered scenarios are under intense pressure from other sources, such as DM direct detection experiments (for example, the Bino/Higgsino candidate, which has a large $\neu{1} \neu{1} h$ coupling, and scatters efficiently off nuclei).

Of course the thermal WIMP framework should be patiently probed in all corners of parameter space. We refer to a selective list of collider studies that have been performed in this regard \cite{Dutta:2011kp}. The high luminosity LHC will certainly shed more light on the viability of thermal DM in the MSSM. 

Given the above, it is important to keep an open mind about conditions of minimality. There are essentially two departures from the minimal scenario that can be considered:  $(i)$ a non-standard cosmological history for DM, such as its origin from the decay of moduli fields in the early universe \cite{MR} or $(ii)$ multi-component DM, in which the cosmological history is thermal, but the relic density is satisfied by the lightest neutralino along with one or more additional candidates, motivated by other physics (not necessarily supersymmetry).

Both these non-minimal options immediately open up vast swathes of parameter space, because the relic density constraint is effectively loosened. The case for non-thermal cosmological histories is particularly strong, given this bias. The reasons are as follows: 

$(a)$ Non-thermal histories can accommodate both the case of thermally under-producing (Wino-like and Higgsino-like) DM as well as the case of thermally over-producing (Bino-like) DM . Note that multi-component DM fails to accommodate the latter. This is important, since it is the Bino-like LSP that one obtains in large parts of parameter space. The correct relic density is obtained in this case through what we will call the ``branching scenario", which works by producing the correct DM number density from a late decaying modulus without relying on further DM annihilation \footnote{We will not consider light $\sim \mathcal{O}(keV)$ DM in this review, for which completely different kinds of non-thermal mechanisms, like freeze-in, are required to get the correct relic density \cite{Queiroz:2014yna}.}. While this fact has been widely known from the original literature in the subject, we feel that it has been somewhat under-appreciated in model building efforts. 

$(b)$ The ubiquity of moduli in string theory, and the various explicit existing models of moduli stabilization, provide a fertile laboratory for exploring the model building challenges of non-thermal histories. 

$(c)$ Both indirect and direct detection DM experiments are beginning to say interesting things about models of non-thermal DM. 

$(d)$ Non-thermal baryogenesis can be achieved with $\mathcal{O}(1)$ couplings of new fields to the MSSM, and the baryon-DM coincidence problem can be addressed by branchings from modulus decay. 


The plan of the paper is as follows. In Section \ref{annbran}, we give the basics of non-thermal DM with particular emphasis on the options available to obtain the correct relic density. In Section \ref{indircol}, we describe aspects of indirect and direct DM as well as collider experiments that have a bearing on non-thermal scenarios. In Section \ref{challenges}, we give the basic challenges of these scenarios, and in Section \ref{UVmodels}, we give some avenues of model building in the modulus sector to address these challenges. In Section \ref{clado}, we describe how the baryon asymmetry-DM coincidence problem may be addressed in this framework. 

Before proceeding, we give an incomplete list of recent works that have investigated non-thermal dark matter in various contexts \cite{alist1}, \cite{alist2}.

\section{The Two Options: Annihilation and Branching} \label{annbran}

In this Section, we discuss the two main scenarios for DM from modulus decay: annihilation (which accommodates candidates like the Wino or the Higgsino) and branching (which accommodates candidates like the Bino).

We consider a scalar field $S$ with mass $m_S$ and decay width 
\be
\Gamma_S = \frac{c}{2\pi} \frac{m^3_S}{M_{\rm P}^2}\,, 
\ee
where $c$ is a model dependent constant.

Assuming that $S$ has acquired a large vacuum expectation value during inflation, it will start oscillating about the minimum of its potential with an initial amplitude $S_0$ when the Hubble expansion rate is $H \sim m_S$. Oscillations of $S$ behave like matter, with an initial energy density $\rho_S = m^2_S S^2_0/2$. The energy density of the universe at this time, dominated by thermal bath, is $\rho_{\rm r} = 3 m^2_S M^2_{\rm P}$.

The quantity $\rho_S/\rho_{\rm r}$ is redshifted $\propto a$, with $a$ being the scale factor of the universe. After using the fact that $H$ is redshifted $\propto a^{-2}$ for a radiation-dominated universe, we find the necessary condition for $S$ to be dominant at the time of decay
\be \label{domination}
{S_0 \over M_{\rm P}} \gg \left({\Gamma_S \over m_S}\right)^{1/4} .
\ee
Decay of $S$ reheats the universe to a temperature:
\be \label{Tr}
T_{\rm r} = c^{1/2} \left(\frac{10.75}{g_*}\right)^{1/4} \left( \frac{m_S}{50\, {\rm TeV}}\right)^{3/2}\, T_{\rm BBN} \,, 
\ee
where $T_{\rm BBN} \simeq 3 ~ {\rm MeV}$ and $g_*$ is the number of relativistic degrees of freedom at $T_{\rm r}$. As a numerical example, for $m_S \sim {\cal O}({\rm TeV})$ and $T_{\rm r} \sim 3$ MeV (in order to be compatible with Big Bang Nucleosynthesis or BBN), Eq. \ref{domination} implies $S$ dominance for $S_0 \gg 10^{13}$ GeV.

If $S$ dominates the universe at a temperature $T_{\rm dom}$, we will have
\bea \label{entropy}
{\rho_{\rm r, after} \over \rho_{\rm r, before}} & = & {T_{\rm dom} \over T_{\rm r}} \, , \nonumber \\
&& \, \nonumber \\
{s_{\rm after} \over s_{\rm before}} & = & \left({T_{\rm dom} \over T_{\rm r}}\right)^{3/4} ,
\eea
where ``before'' and ``after'' are in reference to the epoch of $S$ decay, and we have used the fact that $\rho_{\rm r, after} = \rho_S$.

It is seen from Eq. \ref{entropy} that $S$ decay releases a large entropy that dilutes any pre-existing quantity in the thermal bath. For the above numerical example where $m_S \sim {\cal O}({\rm TeV})$ and $T_{\rm r} \sim 3$ MeV, the entropy release factor can be as large as $10^8$.


Provided that $T_{\rm r} < T_{\rm f} \sim m_\chi/25$, where $T_{\rm f}$ is the thermal freeze-out temperature of the DM candidate $\chi$ with mass $m_{\chi}$, the decay of $S$ will dilute any thermally produced DM by a large factor as mentioned above. However, $S$ decay itself produces DM particles. The abundance of non-thermally produced DM is given by
\be \label{nonthr}
{n_\chi \over s} = {\rm min} ~ \left[Y_S ~ {\rm Br}_\chi ~ , ~ \left({n_\chi \over s}\right)_{\rm thr} ~ \left({T_{\rm f} \over T_{\rm r}}\right) \right] .
\ee
Here 
\be \label{Y}
Y_S \equiv 3 T_{\rm r}/4 m_S 
\ee
is the dilution factor from modulus decay. ${\rm Br}_\chi$ is the branching fraction for production of $R$-parity odd particles from $S$ decay. $(n_\chi/s)_{\rm thr}$ denotes DM abundance obtained via thermal freeze-out that is related to the observed DM relic abundance $(n_\chi/s)_{\rm obs}$ through:
\bea 
\left({n_\chi \over s}\right)_{\rm thr} & = & \left({n_\chi \over s}\right)_{\rm obs} ~ {3 \times 10^{-26} ~ {\rm cm}^3 ~ {\rm s}^{-1} \over \langle \sigma_{\rm ann} v \rangle} \, , \label{thr} \\
& & \,  \nonumber \\
\left({n_\chi \over s}\right)_{\rm obs} & \approx & 5 \times 10^{-10} ~ \left({1 ~ {\rm GeV} \over m_\chi}\right) \, \label{thr2} .
\eea

The abundance of DM particles immediately after their production from $S$ decay is given by $Y_S {\rm Br}_\chi$. If $n_\chi \langle \sigma_{\rm ann} v \rangle < H(T_{\rm r})$, DM annihilation will be inefficient at temperature $T_{\rm r}$. In this case, the final DM relic abundance will be given by the first term inside the brackets in ~Eq. \ref{nonthr}. On the other hand, if $n_\chi \langle \sigma_{\rm ann} v \rangle > H(T_{\rm r})$, annihilation will be efficient right after $S$ decay. This will somewhat reduce the abundance of DM particles produced from $S$ decay, in which case the final relic density will be given by the second term inside the brackets in ~Eq. \ref{nonthr}.

There are therefore two possible scenarios for obtaining the correct DM relic density from $S$ decay:
\begin{itemize}
\item{
{\bf Annihilation Scenario}: If $\langle \sigma_{\rm ann}\rangle > 3 \times 10^{-26} ~ {\rm cm}^3 ~ {\rm s}^{-1}$, then $(n_\chi/s)_{\rm thr} < (n_\chi/s)_{\rm obs}$ (hence ``thermal underproduction''). The large annihilation cross section can reduce the abundance of DM particles produced from $S$ decay to an acceptable level, provided that:
\be \label{anncond}
T_{\rm r} = T_{\rm f} ~ ~ {3 \times 10^{-26} ~ {\rm cm}^3 ~ {\rm s}^{-1} \over \langle \sigma_{\rm ann} v \rangle} .
\ee
The final DM abundance will then be given:
\be
\label{annden}
{n_\chi \over s} = \left({n_\chi \over s}\right)_{\rm thr} ~ {3 \times 10^{-26} ~ {\rm cm}^3 ~ {\rm s}^{-1} \over \langle \sigma_{\rm ann} v \rangle} ~ \left(T_{\rm f} \over T_{\rm r} \right) .
\ee

This scenario can work well in the case of Wino/Higgsino DM, for which $\langle \sigma_{\rm ann} v \rangle > 3 \times 10^{-26} ~ {\rm cm}^3 ~ {\rm s}^{-1}$ as mentioned before, provided that the reheat temperature from $S$ decay satisfies ~Eq. \ref{anncond}.
}
\item{
{\bf Branching Scenario:} If ~Eq. \ref{anncond} is not satisfied, then annihilation will be rendered ineffective. This happens if $T_{\rm r}$ is too low and/or $\langle \sigma_{\rm ann} v \rangle$ is too small.

The first possibility is that $\langle \sigma_{\rm ann} v \rangle > 3 \times 10^{-26} ~ {\rm cm}^3 ~ {\rm s}^{-1}$, but $T_{\rm r}$ is lower than that given in ~Eq. \ref{anncond}. In this case non-thermal Wino/Higgsino DM must be produced via ``Branching Scenario''.

On the other hand, we note that ~Eq. \ref{anncond} can never be satisfied if $\langle \sigma_{\rm ann} v \rangle < 3 \times 10^{-26} ~ {\rm cm}^3 ~ {\rm s}^{-1}$. It is seen from ~Eq. \ref{thr} that this results in $(n_\chi/s)_{\rm thr} > (n_\chi/s)_{\rm obs}$ (hence ``thermal overproduction''). This leaves ``Branching Scenario'' as the only possibility for non-thermal DM production in this case. Bino DM provides a prime example of this case.

The final DM abundance will be the same as that produced from $S$ decay, which follows
\be \label{brden}
{n_\chi \over s} = Y_S ~ {\rm Br}_\chi .
\ee
}
\end{itemize}

\section{Indirect, Direct, and Collider Searches} \label{indircol}

In this Section, we discuss indirect and direct DM detection constraints on non-thermal scenarios. We also discuss avenues for direct probes of Winos and Higgsinos at the LHC.

\subsection{Indirect Detection and Constraints on the Annihilation Scenario}

Fermi data on dwarf spheroidal galaxies and the galactic center already places tight constraints on the annihilation scenario. The limits on annihilation cross section from dwarf galaxies \cite{fermi}, coupled with ~Eq. \ref{anncond},  indicate that $T_{\rm f} \, \lsim \, 100 \, \, T_{\rm r}$ for $m_{\chi} > 40$ GeV and up to a range of a few hundred GeV. This implies $T_{\rm rh} \gsim 100$ MeV for $40 \,\, {\rm GeV} \, \lsim \, m_{\chi} \, \lsim \, \mathcal{O}({\rm few \,\,\, hundred \,\,\, GeV}) $. For $m_{\chi} < 40$ GeV, the Fermi bounds require $\langle \sigma_{\rm ann} v \rangle_{\rm f} < \langle \sigma_{\rm ann} v \rangle_{\rm f}^{\rm th}$,
if DM annihilates into $b {\bar b}$ with $s$-wave domination, implying that the annihilation scenario cannot work in this case.

\begin{figure}[tb]
\includegraphics*[width=0.60\columnwidth]{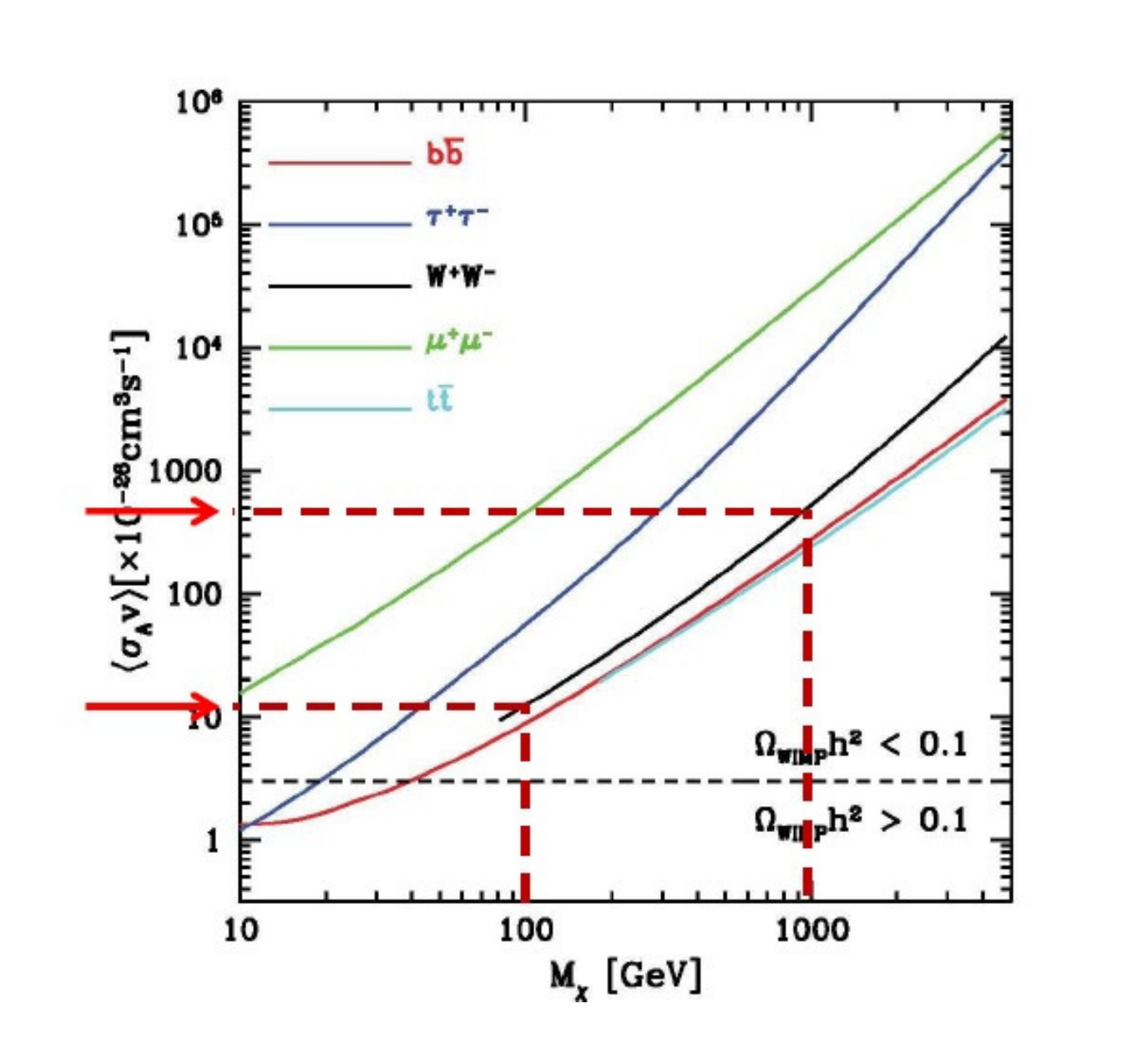}
\vspace*{-.1in}
\caption{Constraints on the annihilation scenario from dwarf spheriodal galaxies. Figure taken from \cite{fermi}. }
\label{Indirect_Detection_Higgsino}
\end{figure}

Fig. \ref{Indirect_Detection_Higgsino} demonstrates the above points. Some representative annihilation cross sections are given below:
\begin{eqnarray} \label{fermi}
&& \langle \sigma_{\rm ann} v \rangle_f \ls 10^{-25} ~ {\rm cm}^3 ~ {\rm s}^{-1} ~ ~ ~ ~ ~ ~ ~ ~ ~ ~ m_\chi = 100 ~ {\rm GeV} \, , \nonumber \\
&& \langle \sigma_{\rm ann} v \rangle_f \ls 3 \times 10^{-24} ~ {\rm cm}^3 ~ {\rm s}^{-1} ~ ~ ~ ~ ~ m_\chi = 1 ~ {\rm TeV} .
\end{eqnarray}

The constraints become stronger when galactic center data is included. According to the analysis in~\cite{Center}, the constraint on the annihilation cross-section from the gamma-ray flux from the galactic center region  is similar for the above neutralino masses to a core of $1$ kpc in the $b \overline{b}$ final states. The constraint on the annihilation cross-section becomes about $4 \times 10^{-26}~ {\rm cm}^3 ~ {\rm s}^{-1}$ for $m_{\tilde\chi} = 100$ GeV for the NFW profile without any core \footnote{We note that an explanation of the PAMELA anomaly requires much larger cross-section \cite{Dutta:2009uf}. However, the explanation of this anomaly can be due to the pulsars. We also note that the bounds on the cross section from dark matter annihilation to neutrinos at the galactic center, obtained by IceCube, are weaker by few orders of magnitude.}.

These kinds of constraints were studied in the case of Higgsino DM, which mainly annihilates into heavy Higgs bosons, $W$ bosons and $t$ quarks via $s$-wave annihilation if $m_\chi$ has necessary phase space for these particles to be produced, in a calculable modulus sector in \cite{Higgsino}. Within the simplest calculable moduli sectors such as KKLT, these bounds on the reheat temperature in the annihilation scenario translate into bounds for the modulus mass from ~Eq. \ref{Tr}:
\begin{eqnarray} \label{td2}
T_{\rm r} &\gsim & 0.4-1.6 ~ {\rm GeV} ~ ~ ~ ~ ~ ~ m_\chi = 100 ~ {\rm GeV}-1 ~ {\rm TeV} \, , \\
m_S &\gsim & \mathcal{O}(1000) ~ {\rm TeV}\,,
\end{eqnarray}
which further translates into a bound on the gravitino mass, and the supersymmetric spectrum, in these models. 

The fact that Wino DM below $\lsim 250$ GeV is in tension with dwarf galaxy data was mentioned in \cite{winodmnonth1}. We refer to the detailed analyses of \cite{winodmnonth2} for the current state of Wino DM. 

As constraints from indirect detection get stronger, it is possible that only over-producing Bino-like candidates may start getting preferred, as is already happening in the low mass $\lsim 40$ GeV region. The branching scenario is the only option for such candidates.


\subsection{Direct Detection, Light DM, and the Branching Scenario}

The branching scenario is strongly preferred as the only option in the mass range $m_{\chi} < 40$ GeV from indirect detection, as we saw above. There are other motivations for this option for light DM candidates coming from direct detection, as we outline below.

The CDMS Collaboration \cite{Agnese:2013dwa} has recently announced results from a blind analysis of data taken with Silicon detectors of the CDMSII experiment in 2006-2007. The collaboration reports DM events that survive cuts with a significance of $3.1 \sigma$ corresponding to DM mass $m_{\rm DM} \sim 8$ GeV and spin-independent scattering cross-section $\sigma_{\rm SI} \sim 10^{-41}$ cm$^{2}$. The excess reported by the CoGeNT collaboration hints at light dark matter in a similar region of parameter space, while CDMS II Ge and EDELWEISS data do not exclude it \cite{cogent}. While XENON100 data would appear to rule out this result at the present time, XENON10 is not that inconsistent with it \cite{cogent}, clearly warranting further probes of this region.

A light DM with cross section in the above range is somewhat challenging for thermal framework. The reasons are as follows. If DM scattering off nuclei occurs through the exchange of new heavy coloured states at the TeV-scale, then obtaining the desired scattering cross-section for a light DM leads to a small annihilation cross section and hence an over-abundance of relic DM at the current epoch assuming standard cosmological evolution. This is precisely the kind of situation suited to the branching scenario.

As an explicit example, we consider the following model for light DM \cite{Allahverdi:2013tca}. We start with the MSSM and introduce new iso-singlet color-triplet superfields $X$ and ${\bar X}$ with respective hypercharges $+4/3$ and $-4/3$, and a singlet superfield $N$ with the following superpotential which is added to the usual MSSM superpotential 
\be \label{superpot}
W_{\rm new}  =  \lambda_{i} X N u^c_{i} + \lambda^\prime_{i j} {\bar X} {d^c_i} d^c_j + M_X X {\bar X} + \frac{M_N}{2} N N \, .
\ee
Here $i,~j$ denote flavor indices (color indices are omitted for simplicity), with $\lambda^\prime_{i j}$ being antisymmetric under $i \leftrightarrow j$. We assume the new colored particles associated with the $X,~\bar{X}$ superfields to have TeV to sub-TeV mass and the scalar partner of singlet $N$, denoted by $\tilde{N}$, will be assumed to have mass in the $8-10$ GeV range and will be identified with the DM particle.

The superpotential coupling $\lambda_i X N u^c_i$ yields an effective interaction between ${\tilde N}_1$ and a quark $\psi$ via $s$-channel exchange of the fermionic component of $X$. The amplitude is given by $i \frac{\vert \lambda_1 \vert ^2}{4M^2_X}({\bar \psi}(k^{\prime}) \gamma^\mu \psi(k))Q_\mu $, where $k_\mu$ is the quark momentum, $p_\mu$ is the momentum of ${\tilde N}_1$, and $Q_\mu = k_\mu+p_\mu$.
%
%
This results in the following spin-independent DM-proton elastic scattering cross section
\be \label{SI}
\sigma^{\rm SI}_{{\tilde N}_1-p} \simeq \frac{\vert \lambda_1 \vert^4}{16 \pi} \frac{m^2_p}{M^4_X},
\ee
where $m_p$ is the proton mass. It is seen that for $\vert \lambda_1 \vert \sim 1$ and $M_X \sim 1$ TeV, which is compatible with the LHC bounds on new colored fields, we get $\sigma^{\rm SI}_{{\tilde N}_1-p} \sim {\cal O}(10^{-41})$ cm$^2$. We note that this scenario easily evades bounds coming from monojet searches at colliders. The pair production of fermionic components of $X,~{\bar X}$ superfields, which are $R$-parity odd, will produce 4 jets plus missing energy final states at the LHC in this model. In the non-supersymmetric version of the model, where $N$ fermion is the DM candidate, the absence of $R$-parity fields results in  missing energy final states with 2 and 3 jets only, which will allow us to distinguish the two scenarios.

The superpotential coupling $\lambda_i X N u^c_i$ also results in annihilation of ${\tilde N}_1$ quanta into a pair of a right-handed quark and left-handed antiquark of the up-type. Considering that $m_{{\tilde N}_1} \sim {\cal O}(10~{\rm GeV})$, only annihilation to up and charm quarks is possible when temperature of the universe is below $m_{{\tilde N}_1}$. The annihilation rate is given by
\be \label{ann}
\langle \sigma_{\rm ann} v_{\rm rel} \rangle \simeq \frac{\vert \lambda_1 \vert^4 + \vert \lambda_2 \vert^4 +2\vert\lambda_1\lambda^*_2\vert^2}{8 \pi} \frac{\vert {\vec p} \vert^2}{M^4_X} ,
\ee
where ${\vec p}$ is the momentum of annihilating ${\tilde N}_1$ particles. It is seen that for $\vert \lambda_1 \vert \sim \vert \lambda_2 \vert \sim 1$, $m_{{\tilde N}_1} \sim {\cal O}(10 ~ {\rm GeV})$, $M_X \sim 1$ TeV we have $\langle \sigma_{\rm ann} v_{\rm rel} \rangle_{\rm thermal} \ll 3 \times 10^{-26}$ cm$^3$ s$^{-1}$. Therefore thermal freeze-out yields an over-abundance of ${\tilde N}_1$ particles.

This implies that obtaining the correct DM relic density requires a non-thermal scenario which can address thermal over-abundance. This is precisely the branching scenario.

One can write down operators that can plausibly lead to the branching scenario (specific UV settings will be discussed in Section \ref{UVmodels}). The main requirement, as described in Section \ref{rodd}, is that modulus decay to $R$-odd states be suppressed to $\mathcal{O}(10^{-3})$. $S$ can mainly decay into scalar components of $X,~{\bar X}$ superfields (denoted by ${\tilde X},~{\tilde {\bar X}}$ respectively), which are $R$-parity even fields, through a coupling $K \supset \lambda_X S^\dagger  X {\bar X}$ in the K\"ahler potential.
The decay into the $R$-parity even fermions suffers chiral suppression. The decays of $S$ to $R$-parity odd gauginos can be suppressed by suitable geometric criteria e.g., by constructing the visible sector at a singularity and selecting $S$ to be the volume modulus in large volume compactification scenarios \cite{Allahverdi:2013noa} (we note here that tree level modulus coupling through the gauge kinetic function does \textit{not} suppress gaugino production -  a fact overlooked in some of the early literature). The decay of $S$ to other $R$-parity odd MSSM fields like squarks and sleptons is suppressed after using the equations of motion. The decay to the gravitino can also be kinematically suppressed for superheavy ($\sim 10^{12}$ GeV) gravitinos. Finally, the decay of $S$ to ${\tilde N}_{1,2}$ is suppressed by preventing the K\"ahler potential coupling $\lambda_N S^\dagger {\tilde N}^2$ with symmetries.

\subsection{Collider Searches for Winos and Higgsinos: Vector Boson Fusion and Charged Track Analysis}

Observing candidates such as the non-thermal Wino or the Higgsino at the LHC remains an important complementary objective. The main challenge to a direct probe of the electroweak sector at the LHC is the small production cross section of neutralinos, charginos, and sleptons compared to the coloured sector. 

Vector boson fusion (VBF) processes, characterized by two jets with large dijet invariant mass in the forward region and in opposite hemispheres, are a promising avenue to search for new physics. Two recent studies have used VBF processes to investigate the chargino/neutralino sector of supersymmetric theories (\cite{Delannoy:2013ata}, \cite{Dutta:2012xe}). Direct production of  the lightest neutralino by VBF processes in events with two forward jets and missing transverse energy in the final state was studied at the 14 TeV LHC, providing a search strategy also applicable to compressed spectra that can be carried out with an experimentally plausible and bias-free trigger. A simultaneous fit of the signal rate and missing energy distribution was used to measure the mass and composition of the lightest neutralino as well as the dark matter relic density. It was shown that Wino (Higgsino) masses can be probed (at 5$\sigma$) up to approximately 600 (400) GeV with a luminosity of 1000 fb$^{-1}$ at LHC14. In a separate study, Wino/chargino production by VBF processes followed by decay to the lightest neutralino through either a light slepton or a light stau was studied in final states with two forward jets, missing energy, and either dilepton or ditau, respectively. 

The mass reaches of these theoretical studies should be tempered by the fact that high pileup (PU) conditions are generally expected to degrade them considerably. We refer to \cite{cmspileup} for the CMS discovery reaches at $3000$ fb$^{-1}$ with PU $ = 140$.

A totally complementary method of probing Winos is the ATLAS charged track analysis \cite{TheATLAScollaboration:2013bia}. With $20$ fb$^{-1}$ of data at LHC8, the exclusion bounds on Winos reach approximately $\sim 250$ GeV, when matched with a two-loop theoretical calculation of the mass splitting between the charged and neutral Wino \cite{Ibe:2012sx}.


\section{Outstanding Challenges} \label{challenges}

In this Section, we describe the main challenges of obtaining viable non-thermal scenarios.

\subsection{Moduli-induced Gravitino Problem}

This is a classic problem in most non-thermal scenarios \cite{gravProbl}. The gravitino must decay before BBN; moreover it is important to note that the gravitino is typically the last particle to decay, not the modulus. A source of non-thermal DM production, in addition to that coming from $S$ that has been studied already, is now provided by the gravitino. Since the gravitino decays at a temperature $\ll {\cal O}({\rm GeV})$, and DM annihilation rate must satisfy the Fermi bounds~(\ref{fermi}), annihilation is very inefficient at this time. Thus, the density of LSPs produced from gravitinos is the same as the density of gravitinos (since $R$-parity is conserved). Therefore, we require
\begin{eqnarray} \label{gravdens}
{n_{3/2} \over s} \ls 5 \times 10^{-10} ~ \left({1 ~ {\rm GeV} \over m_\chi}\right) \, .
\end{eqnarray}

Now, the density of gravitinos is in turn set by their production via thermal and non-thermal processes in the early universe. Modulus decay dilutes gravitinos that were produced in the prior epochs (e.g., during inflationary reheating) by a huge factor. Thermal gravitino production after modulus decay is highly suppressed due to the low decay temperature $T_{\rm r} \sim {\cal O}({\rm GeV})$.

Therefore, the  density of gravitinos is given by what is produced directly from modulus decay $S \rightarrow {\tilde G} {\tilde G}$, and it is $(n_{3/2}/s) = {\rm Br}_{3/2} (3 T_{\rm r}/4 m_S)$, where ${\rm Br}_{3/2}$ is the branching ratio for $S \rightarrow {\tilde G} {\tilde G}$ process. We then find
\begin{eqnarray} \label{gravbr}
{n_{3/2} \over s} \sim 5 \times 10^{-8} ~ \left(\frac{m_S}{100\, {\rm TeV}}\right)^{1/2} ~ {\rm Br}_{3/2} \, .
\end{eqnarray}

For typical values of $m_S$  and $100~{\rm GeV} \leq m_\chi \leq 1$ TeV, using ~Eq. \ref{gravdens},~Eq. \ref{gravbr} one gets the following absolute upper bound:
\be \label{brconst}
{\rm Br}_{3/2} \ls 10^{-5} .
\ee
\\
\\
\noindent
\textbf{Summary}: \textit{Any successful scenario for non-thermal DM production from modulus decay must suppress gravitino production to $\sim \mathcal{O}(10^{-5})$, or decouple the gravitinos kinematically.}

\subsection{The Branching Scenario Requires Suppression of 2-Body Decay of Modulus to $R$-odd States.} \label{rodd}

The branching scenario is summarised by the requirements of ~Eq. \ref{thr2} and ~Eq. \ref{brden}, as well as the expression for the reheat temperature, ~Eq. \ref{Tr}. For $m_S \sim 1000$ TeV, one obtains $T_r \sim 100$ MeV, and thus $Y_S \gsim 10^{-8}$. This shows that ${\rm Br}_\chi$ cannot be $\mathcal{O}(1)$, if one is to obtain the correct relic density.

For the typical range of modulus mass of $(100-1000)$ TeV, with $c \sim 0.1-1$, the desired dark matter abundance is obtained for
\be
{\rm Br}_\chi \lsim 10^{-3} ~ ~ ~, ~ ~ ~ 5 ~ {\rm GeV} \lsim m_\chi \lsim 500 ~ {\rm GeV}.
\ee
This requires the two-body decays of the modulus to $R$-parity odd particles to be suppressed. Note, however, that these particles are inevitably produced from three-body decays of the modulus. It is striking that the three-body decays are suppressed by a factor of $\sim 10^{-3}$ relative to two-body decays based on the phase space factors.

One can, of course, loosen the condition  $c \sim 0.1-1$ and make $c \ll 1$. This will make $T_r$ smaller for the same $m_S$, lowering $Y_S$ and enabling ${\rm Br}_\chi$ to evade the upper bound.
\\
\\
\noindent
\textbf{Summary}: \textit{the branching scenario for non-thermal DM requires the two-body decays of the modulus to $R$-parity odd particles to be suppressed (this can be traded with suppressed overall decay $c \ll 1$). We note that tree level couplings of the modulus to the visible sector through the gauge kinetic function violates this condition, because of equal decay to gauginos and gauge bosons.}

\subsection{Correlation with Dark Radiation}

The moduli are gauge singlets and so
they do not necessarily prefer to decay into visible sector fields. Light axion-like particles generally exist in perturbatively stabilized string compactifications in type IIB; while they may be eaten up by anomalous $U(1)$ symmetries, at least one such axion may be argued to survive \cite{Allahverdi:2014ppa}, \cite{Queiroz:2014ara}. Thus,
since light hidden sector degrees of freedom like axion-like particles can be expected to exist,
the branching ratio into them could be non-negligible, so giving a number of
effective relativistic species which is above the tight bounds from cosmological observations,
$\Delta N_{\rm eff}\simeq 0.5$. By combining present upper bounds on $\Delta N_{\rm eff}$ with lower bounds on the reheating temperature as a function of the dark matter mass from Fermi data, one can obtain strong constraints on the ($\Delta N_{\rm eff}$,$m_{DM}$)-plane. Most of the allowed region in this plane corresponds to non-thermal scenarios with Higgsino-like dark matter \cite{Allahverdi:2014ppa}. Thermal dark matter strongly prefers the Standard Model value of $N_{\rm eff}$. Future CMB polarization and Large Scale Structure experiments could in fact measure $\Delta N_{\rm eff}$ down to values which would be able to definitively discriminate between a thermal and non-thermal history for DM.
\\
\\
\noindent
\textbf{Summary}: \textit{one has to be careful not to violate current bounds and over-produce dark radiation. In fact, the value of $\Delta \, N_{{\rm eff}}$ will generally have consequences for which scenario of non-thermal DM (annihilation or branching) is preferred.}


\subsection{``Genericness" Problem}

The reheat temperature of the modulus has to be below the freeze-out temperature of DM (to be interesting in the first place), and above the Big Bang Nucleosynthesis (BBN) bound of $\sim \, \mathcal{O}(3)$ MeV (in order not to ruin the successes of BBN). Typically, this implies, from ~Eq. \ref{Y} and  ~Eq. \ref{Tr}, that the mass of the modulus should be in a window between $\mathcal{O}(10)$ TeV to $\mathcal{O}(\, {\rm few \,\,\,}1000)$ TeV, for typical values of its overall decay pre-factor $c$. The question of whether such moduli exist generically in string compactifications is unclear. Certainly, in the most widely studied examples, there are such moduli, as we will elaborate on in the next section. There have also been arguments at the level of effective supergravity pointing to the existence of moduli with masses close to the gravitino mass \cite{Acharya:2010af}.
\\
\\
\noindent
\textbf{Summary}: \textit{interesting and viable non-thermal DM physics requires late-decaying moduli with mass between $\mathcal{O}(10)$ TeV to $\mathcal{O}(\, {\rm few \,\,\,}1000)$ TeV. Whether such moduli exist generically in string compactifications is unclear.}

\section{UV Models: Moduli Sector} \label{UVmodels}

In this Section, we describe some specific UV models for the modulus sector.

\subsection{KKLT}

The essential elements in a KKLT-type model are: $(1)$ background fluxes on a type IIB Calabi-Yau three fold giving a Gukov-Vafa-Witten superpotential contribution that fixes complex structure moduli, and $(2)$ gaugino condensation on $D7$ branes or Euclidean $D3$ instantons giving a non-perturbative superpotential contribution that fixes the Kahler moduli. An additional contribution to the scalar potential coming from anti-${D}3$ branes then lifts the solution to a de Sitter vacuum. 

The late-decaying modulus is the volume modulus, which has mass in the correct range. The decay modes are as follows: 

$(i)$ The gauginos and gauge bosons couple through the gauge kinetic function. The modulus decays with \textit{equal branching ratio} to gauge bosons and gauginos, and these are the primary decay modes of the modulus. $(ii)$ Visible sector fermions and scalars couple to the modulus through the Kahler potential and soft terms. The modes suffer chiral suppression, except for decays to the Higgses, through a Giudice-Masiero type coupling. $(iii)$ The gravitino couples to the modulus and the branching to gravitino is $\sim \, 1\%$.
\\
\\
\noindent
\textbf{Evaluation:} \textit{the model has a modulus in the required mass range. It fails the gravitino problem and has the incorrect branching fractions for the branching scenario to work. There is no problem with dark radiation since the volume axion is heavy and kinematically decoupled.}

\subsection{LVS}

Another well-studied moduli stabilisation mechanism in type IIB string theory is
the LARGE Volume Scenario. We summarise some recent work on non-thermal DM in this context \cite{Allahverdi:2013noa}.

In this framework,
all the moduli are fixed by background fluxes, D-terms from anomalous $U(1)$s,
and the interplay of non-perturbative and $\alpha'$ effects.
The simplest realisation involves an internal volume of the form:
\be
\mathcal{V} = \tau_{\rm big}^{3/2}-\tau_{\rm np}^{3/2}-\tau_{\rm inf}^{3/2}-\tau_{\rm vs}^{3/2}\,,
\ee
where the $\tau$'s are K\"ahler moduli parameterising the size of internal 4-cycles.
The visible sector (a chiral MSSM- or GUT-like theory)
is built via space-time filling D3-branes sitting at the singularity
obtained by shrinking $\tau_{\rm vs}$ to zero size by D-terms.
%
%
%
%
%
All the relevant energy scales in the model are set by value of $\mathcal{V}$. 
%

%
One typically obtains $M_{\rm GUT} \simeq 10^{16}$ GeV,
$m_{3/2}\sim 10^{10}$ GeV, $m_{\tau_{\rm big}} \simeq 5 \times 10^6$ GeV
and $M_{\rm soft} \simeq 1$ TeV.

The modulus $\tau_{\rm big}$ has mass in the correct window and can lead to non-thermal DM (the other Kahler moduli are much heavier). The gravitino is extremely heavy and kinematically decoupled, while the scenario still has TeV-scale supersymmetry. The leading decay channels for $\tau_{\rm big}$ are to Higgses (through a GM term) and closed string axions. In particular, in contrast to the KKLT scenario, the volume modulus only has a radiatively induced coupling to the gauge kinetic function; thus, decays to gauginos (and gauge bosons) is loop suppressed in this case.

Depending on the way in which the modulus couples to the visible sector, there are two regimes
for the reheat temperature $T_{\rm r}$. The case of high $T_{\rm r}\simeq 1$ GeV
is realised when the modulus decays mainly to Higgses, and corresponds to the annihilation scenario. The branching scenario may be accommodated and the reheat temperature lowered to $T_{\rm r}\simeq 10$ MeV if the modulus decays mainly to gauge bosons (or if the decay to Higgses is suppressed).
Axionic dark radiation overproduction may be avoided either by the presence of anomalous $U(1)$s which
eat dangerous axions or by allowing suitable couplings in the Giudice-Masiero term.
%
\\
\\
\noindent
\textbf{Evaluation:} \textit{the model has a modulus in the required mass range. It avoids the gravitino problem through kinematic decoupling (but still achieves a TeV-scale SUSY spectrum). Both annihilation and branching scenarios may be accommodated by suitable choice of Kahler potential couplings. The overproduction of dark radiation may be avoided if bulk and local axions are eaten by anomalous $U(1)$s, or by suitable choice of parameters.}

\subsection{Visible Sector Modulus}

To address the various challenges of non-thermal scenarios with a gravitationally coupled modulus, one needs to satisfy non-trivial conditions on the K\"ahler geometry of the underlying effective supergravity theory, as we have seen. An alternative is to shift model-building to a visible sector scalar field that decays and gives rise to the non-thermal history.

In a model constructed in \cite{Allahverdi:2012gk}, the late decay of an $R$-parity even scalar field $S$ that is a SM singlet (but may be charged under a higher rank gauge group) produces DM. $S$ is coupled to new colored fields $X,~{\bar X}$ with a mass relation $m_S \ll m_X$. 
The superpotential of the visible sector is $W_{\rm visible} = W_{\rm MSSM} + W_{N,X} + W_{S}$, where $W_{N,X}$ is given by ~Eq. \ref{superpot} and
%
%
%
\be\label{superpot2}
W_{S} = h S  X {\bar X} + \frac{1}{2} M_S S^2 \,\, .
\ee

Note that we will the superpotential in ~Eq. \ref{superpot} as a sort of template in a number of different examples here and later as well, with vastly different mass and parameter selections.

 Assuming that all $R$-parity odd colored fields are heavier than $S$, it will dominantly decay into gluons at the one-loop level. The combination of a small coupling $h$ between $S$ and $X,~{\bar X}$, the mass relation $m_X \gg m_S$, and the one-loop factor can lead to a late decay of $S$ that yields a low reheat temperature $T_{\rm r}$. One can obtain $T_{\rm r} \sim {\cal O}({\rm GeV})$ for $m_S \sim 1$ TeV, $m_X \sim 50$ TeV, and $h \sim 10^{-6}$.

The branching fraction for $S$ decay to DM particles ${\rm Br}_\chi$ depends on the nature of the LSP. If the LSP is Bino, then $S$ can decay to a pair of DM particles at the one-loop level, where ${\rm Br}_\chi \gsim 10^{-6}$ for $m_\chi \geq 60$ GeV. For a Higgsinos LSP, the same decay occurs at the two-loop level yielding ${\rm Br}_\chi \sim 10^{-10}-10^{-5}$ (the exact value depending on the model parameters) for $m_\chi \sim 100-500$ GeV. As a consequence, one can obtain the observed relic abundance in both larger and smaller annihilation cross-section regions of SUSY parameter space.

Gravitino production from $S$ decay is naturally suppressed by the virtue of $S$ belonging to the visible sector. Moreover, ${\rm Br}_\chi$ can be made sufficiently small by choosing the model parameters, which is essential for a successful realization of the branching scenario.
\\
\\
\noindent
\textbf{Evaluation:} \textit{late decay of a visible sector scalar can avoid the gravitino problem. It can accommodate both branching and annihilation scenarios through a combination of charge selection, interactions, and kinematics. Late decay typically will require a small Yukawa coupling $\sim \, \mathcal{O}(10^{-6})$.}

\section{Cladogenesis: The Baryon-DM Coincidence Problem} \label{clado}

The fact that the yield from modulus decay $Y_S$ is a small number $\gsim 10^{-8}$ is suggestive. It is small because the reheat temperature for a modulus with mass $\sim 1000$ TeV is so low (due to the fact that it decays gravitationally). Given that in baryogenesis and DM physics we are trying to explain number densities $\sim \, \mathcal{O}(10^{-10})$, it is interesting to think that this value is driven by the smallness of $Y_S$, with the remainder being accounted for by branching ratios and loop factors \cite{Allahverdi:2010rh}, \cite{Allahverdi:2010im}. 

One can consider DM and baryon asymmetry both being directly produced from a common source, i.e., the decay of a modulus (hence the name Cladogenesis). In particular, DM annihilation is irrelevant i.e. we are in the branching scenario. We note that no asymmetry is required in the DM sector (these models do not fall within the class of asymmetric DM models).

One then has the following for the baryon and dark matter density ratio:
\be
\frac{\Omega_{\rm B}}{\Omega_{\rm DM}} \, \simeq \,
\frac{1 ~ {\rm GeV}}{m_\chi} \times \frac{\epsilon \, {\rm Br}_N}{{\rm Br}_\chi} \,\,.
\ee
In the above, $N$ is a species whose decay gives rise to baryon asymmetry, with $\epsilon$ being the asymmetry per decay. The branching scenario requires ${\rm Br}_\chi \lsim 10^{-3}$ and the asymmetry per decay is at least loop suppressed for natural $\mathcal{O}(1)$ couplings leading to $\epsilon \, \sim \, \mathcal{O}(10^{-1})$. The value of ${\rm Br}_N$ is model-dependent, but if $N$ is non-colored, simple counting of degrees of freedom suggests ${\rm Br}_N \, \sim \mathcal{O}(10^{-2})$. One then obtains $m_\chi \, \sim \, \mathcal{O}(10)$ GeV.

As an explicit model, one can work out a variant of the superpotential in ~Eq. \ref{superpot}, but with two flavors of singlets $N$, and which can now be arranged to be heavy in this case. The interference between the tree-level and one-loop diagrams in the $N_\alpha \rightarrow X u^c$ decay generates a baryon asymmetry. For $\mathcal{O}(1)$ phases and couplings (which are allowed by experimental bounds), the asymmetry per decay is given by
\be
\epsilon \sim \frac{1}{8 \pi} \frac{[{\rm Tr} (\lambda \lambda^{\dagger})] ~ [{\rm Tr} (\lambda^{\prime} \lambda^{\prime \dagger})]}{{\rm Tr} (\lambda \lambda^{\dagger})} \sim 0.1 \,\,.
\ee
This, along with $Br_{N} \sim 10^{-2}$, yields the correct baryon asymmetry for $Y_S \sim 10^{-7}$.
\\
\\
\noindent
\textbf{Summary}: \textit{The dilution factor from modulus decay is  $\sim 10^{-8}$. Don't throw away a small number you got for free.}

\section{Acknowledgement}

I would like to thank Rouzbeh Allahverdi, Michele Cicoli, Bhaskar Dutta, Rabi Mohapatra, and Scott Watson for collaboration and discussions. I would also like to thank the organizers of CETUP$^*$ and PPC2013 where some of this work was presented. This work is supported by NASA Astrophysics Theory Grant NNH12ZDA001N.

\end{document}